# High efficiency, high quality factor active membrane metasurfaces with extended Kerker effect

Junxing Fan[1,#], Ye Zhou[2,#], Zhanqiang Xue[1], Guizhen Xu[1], Junliang Chen[1], Hongyang Xing[1], Longqing Cong[1, 3, *]

[1]State Key Laboratory of Optical Fiber and Cable Manufacture Technology, Department of Electrical and Electronic Engineering, Southern University of Science and Technology, Shenzhen 518055, China

[2]China-UK Low Carbon College, Shanghai Jiao Tong University, Shanghai 200240, P. R. China

[3]Guangdong Key Laboratory of Integrated Optoelectronics Intellisense, Southern University of Science and Technology, Shenzhen 518055, China

[#]These authors contributed equally to the work

[*]Email: conglq@sustech.edu.cn

**Abstract:** Efficient, low-power consumption, and highly integrated optoelectronic devices remain a critical yet challenging goal. Here, we propose a paradigm of broadband Kerker effect that synergizes Kerker's condition with bound states in the continuum (BICs) to overcome these challenges. By extending Kerker's condition through dual BIC dispersion engineering, we demonstrate a high-efficiency beam deflector using a membrane metasurface. This strategy simultaneously achieves robust parameter tolerance and sensitive narrow linewidth resonances, which are otherwise contradictory. The membrane metasurface based on broadband Kerker effect achieves an absolute beam deflection efficiency exceeding 92%, with exceptional spectral and spatial selectivity characterized by a 4 GHz linewidth, a 2.8° divergence angle, and a quality factor of 114 in experiments. Furthermore, the membrane metasurface reveals a 94% transmission intensity modulation at a pump intensity as low as 0.5 W/cm$^2$. The extended broadband Kerker effect establishes a scalable paradigm for energy-efficient and integrable optoelectronic devices, paving the way for transformative advancements in next-generation wireless communications and Lidar.

## Introduction

The burgeoning field of metasurfaces offers innovative solutions for manipulating electromagnetic waves at high frequencies and is emerging as a powerful platform for optical applications, including compressed imaging[1], biosensors[2], and detectors[3]. By introducing an in-plane wavevector through distributed propagating phase or geometric phase (Pancharatnam-Berry or P-B phase), metasurfaces enable flexible electromagnetic wavefront manipulation — a crucial capability for non-line-of-sight transmission in next-generation wireless communications[4, 5, 6]. This passive beam-steering strategy, underpinned by generalized Snell's law, has garnered substantial research attention[7]. Although notable progress has been made since the pioneering work on metasurfaces, beam-deflection efficiency has consistently been a major obstacle for practical applications. For instance, the initial V-shaped plasmonic nanoantenna exhibited a theoretical efficiency upper limit of 25% due to the existence of reflection and transmission channels and the polarization conversion limit. Subsequent solutions have focused on reducing the number of radiation channels and improving polarization conversion efficiency. Advances such as reflective metasurfaces with a metal–insulator–metal configuration have improved efficiency beyond 80% in metasurface holograms by eliminating transmission channel[8]. In transmission mode, multilayer metasurfaces, e.g., bilayer designs with coupled electric and magnetic dipoles or constructing a Fabry-Perot cavity, realized transparent metasurfaces with efficiency up to 86%[9]. However, the complexity of multilayer architectures and intrinsic metallic losses hinder further scalability at higher frequencies.

Dielectric metasurfaces provide an attractive alternative, combining scalability, simplified design, and ease of fabrication supported by Mie resonances[10]. For instance, metasurfaces based on Kerker's condition leverage degenerate electric and magnetic dipoles to achieve unidirectional transmission, theoretically reaching 100% efficiency at the crossing point of dispersion curves (Fig. 1A). However, efficiency sharply diminishes away from the crossing point, as constructing a phase gradient requires a diverse unit cell parameter (Fig. 1B). As a result, the measured beam deflection efficiency falls far short of expectation under the conventional single-point Kerker's condition. Furthermore, broad spectral resonances with low quality factors ($Q$) below 20 limit selectivity, necessitating higher tuning thresholds and greater energy consumption for dynamical beam deflection (Fig. 1C).

In comparison, nonlocal metasurfaces possess the capability to mold wavefront profiles with exceptional spectral and spatial selectivity[11] but remain limited in efficiency[12]. One solution to improve efficiency involves breaking the out-of-plane symmetry for perfect chiral excitation, albeit at the expense of fabrication complexity and difficulty[13]. Notably, achieving simultaneous efficient control over wavefronts with resonance phase and high quality factors remains an outstanding challenge. As a typical nonlocal mode, bound states in the continuum (BICs)[14, 15], a class of nonradiative states within the continuous spectrum, have emerged as a prominent platform for studying high quality factor wavefront molding. BICs are considered a powerful mechanism for enhancing light-matter interactions and have been applied in fields such as lasers[16, 17, 18], sensing[19, 20, 21, 22, 23], modulators[11, 24, 25, 26, 27], unidirectional radiation[28, 29, 30, 31, 32], and high harmonic generation[33, 34, 35]. Wavefront manipulation based on q-BICs has recently emerged as a major research direction. By combining grating diffraction with ultra-high-$Q$ resonances, prior studies have demonstrated highly frequency-selective beam deflection functionalities[36, 37, 38, 39]. The Kerker effect has proven particularly effective for directional radiation control [40], inspiring various efforts to integrate Kerker physics with q-BIC based resonances [41]. Notably, ultra-high-$Q$ metasurfaces incorporating active materials offer remarkable tunability, requiring only minimal refractive index modulation to achieve effective wavefront control [42]. Experimental demonstrations of reflective and transmissive devices have shown that combining Kerker effects with high-$Q$ resonances can exceed the 25% theoretical efficiency limit of nonlocal metasurfaces, reaching up to 55.6% [43, 44, 45]. However, the issue of low transmittance at non-degenerate points in conventional single-point Kerker configurations remains a critical challenge.

Here, we fuse Kerker's condition and BIC physics to achieve high-efficiency and high quality factor (high-$Q$) wavefront deflection in transmission mode using a free-standing, membrane metasurface. By manipulating the dual-mode dispersion, we extend Kerker effect into the broadband regime in momentum space (Fig. 1D), reconciling the otherwise contradictory between robust parameter tolerance and sensitive narrow linewidth resonances. This approach enables stable high efficiency across all metasurface unit cells (Fig. 1E), yielding an exceptional absolute efficiency exceeding 92%. The simultaneous high-$Q$ resonances enhance the spectral and spatial selectivity of the deflected beam while reducing the tuning threshold to modify the deflected beam for active applications (Fig. 1F). We experimentally demonstrated this paradigm using a free-standing silicon membrane metasurface with a thickness of 115 μm in terahertz regime, which is compatible with integration into sources for applications in next-generation wireless communications and the Internet of Things.

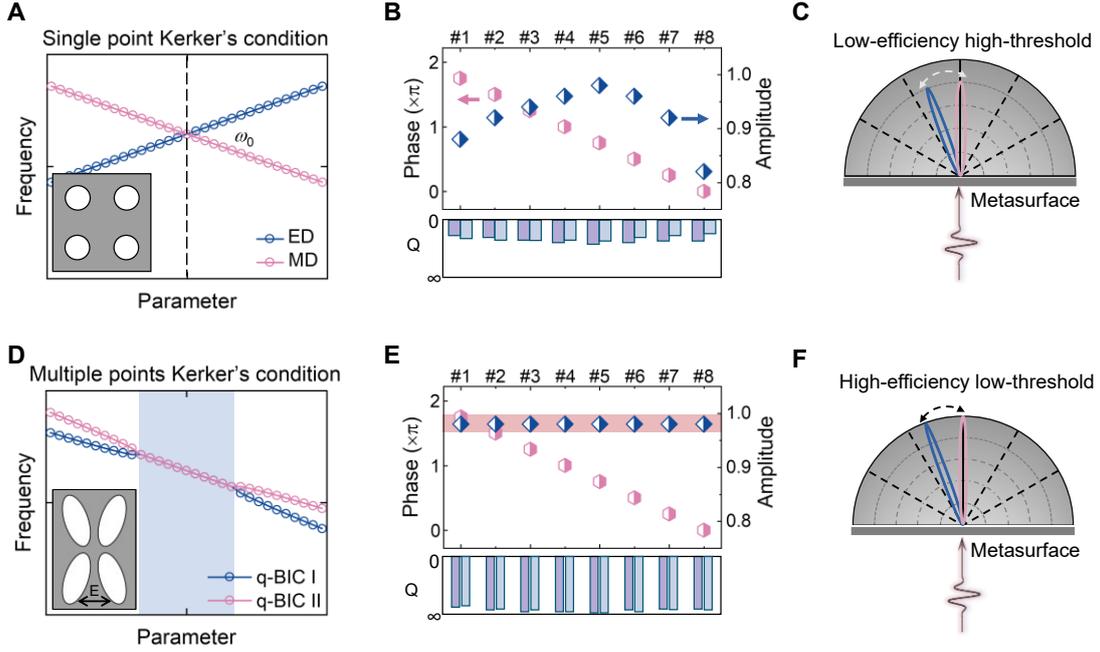

**Fig. 1 Conventional and extended Kerker effect.** (**A**), (**D**) Schematic diagrams showing the band diagrams of conventional single-point Kerker's condition and extended Kerker effect. Insets depict the geometry of the unit cells. (**B**), (**E**) Searching for eight unit cells with phase gradient to form a supercell under the single-point and extended Kerker's conditions. The transmission amplitude of each unit cell varies to around 0.8 under conventional condition while all the unit cells reserve ideal amplitude across all the unit cells under the extended Kerker's condition. By employing BICs, the quality factors of each unit cell can be tuned to a much larger value than conventional dipoles. (**C**), (**F**) Schematic illustration showing the beamform quality and tuning threshold under the two different conditions. Here, the threshold refers to the minimum intensity of external stimulus required to switch the deflected beam back to normal radiation.

## Results

### Extended mode degeneracy in momentum space

We began by examining the mode properties in momentum space using a rectangular lattice comprising diatomic resonators, specifically an array of elliptical holes in a silicon membrane (Fig. 2A). The two modes exhibit opposite dispersion near the $\Gamma$ point, which poses challenges for achieving extended degeneracy. A closer examination of the band diagram reveals that the dispersions flatten near the $Y$ point, displaying similar slopes as they evolve in momentum space (Fig. 2B). However, these guided modes, being below the light line, prevent coupling to far-field radiation, and cannot be applied in wavefront engineering. This limitation can be resolved by doubling the lattice period along the $y$ axis. The corresponding Brillouin zone folding shifts the modes at $Y$ point in the rectangular lattice to $\Gamma$ point, enabling coupling to free space[46]. The real and momentum space correspondence during band-folding process is schematically shown in Fig. 2A with an artificially doubling of lattice along the $y$ axis. These modes at $Y$ point (0.521 THz, 0.501THz), originally bounded, are artificially folded within the continuum forming bound states in the continuum (BIC) with their quality factor ($Q$) dynamics characterized in detail as shown in Fig. 2C[24, 47]. Figure 2D shows the field distributions of the eigenmodes at the $\Gamma$ point, where the mode symmetry—characterized by the $A_2$-type irreducible representation of the $C_{2v}$ point group—prevents coupling to free space under $C_2$ symmetry (see supplementary information 2 for details).

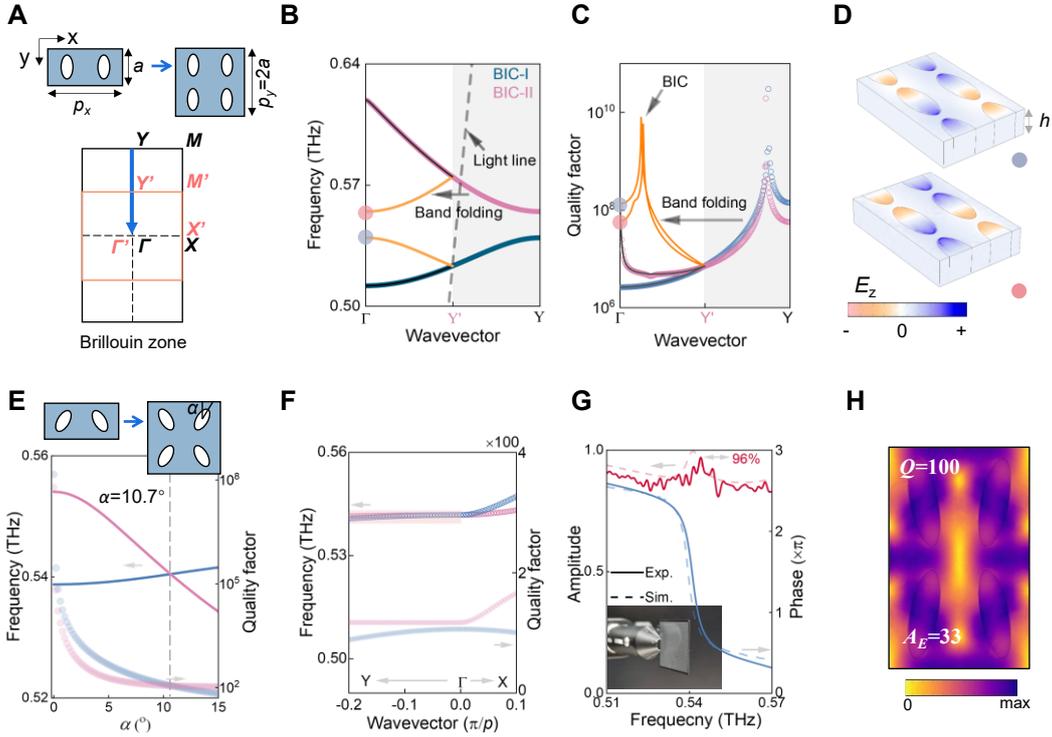

**Fig. 2 Band folding to access extended Kerker effect.** (**A**) Correspondence between real and momentum space. Brillouin zone and modes are artificially folded by doubling the period of lattice along the $y$ axis. (**B**) Artificial band structure evolution when the unit cell period is doubled along the $y$ axis. The guided modes folded from the $Y$ point to the $\Gamma$ point are indicated in the figure by pink (0.521 THz) and blue (0.501THz) dots, respectively. (**C**) Evolution of the corresponding quality factors. We note that values of quality factors at $\Gamma$ points for the orange lines are limited by simulation accuracy and convergence in the 3D solvers. (**D**) Electric field distributions of the two folded BICs in the tetratomic unit cell. (**E**) Parameter searching for the degenerate point by tuning asymmetry angle $\alpha$ in the symmetry-broken tetratomic unit cell and crossing occurs at $\alpha = 10.7°$ where the quality factors are also coincident. (**F**) Extended degenerate BICs along $\Gamma$ to $Y$ point with the symmetry-broken tetratomic unit cell after parameter optimization. The corresponding quality

factors are similar in values. (**G**) Measured and simulated transmission amplitude and phase of sample with the optimized geometries. The inset shows the optical image of the free-standing thin-film sample. The dashed and solid lines represent simulated and measured data, respectively. (**H**) Electric field distributions of the optimized unit cell with the largest local field enhancement factor of 33.

Far-field radiation leakage can be controlled by lifting the symmetry, characterized by the asymmetry degree ($\alpha$) in the tetratomic unit cell (Fig. 2E). The proposed membrane metasurface is a patterned free-standing silicon ($\varepsilon = 11.7$) film with elliptical air holes suspended in air to minimize mode coupling. By adjusting unit cell parameters, we shift the frequencies of the folded and flattened modes, achieving a crossing in both frequency (0.542 THz) and $Q$ factors at $\alpha = 10.7°$ ($Q = 100$, Fig. 2F). As predicted by the band diagram, the two modes exhibit similar dispersion, maintaining degeneracy across a broad range of wavevector along the $y$ axis ($k_y$), as validated in Fig. 2E. The corresponding $Q$ factors also exhibit minimal divergence, ensuring the satisfaction of Kerker's condition across a large range of the parameter space. The optimized geometries of the unit cell are as follows: a thickness of $h = 116.2$ μm, long and short axes of elliptical holes measuring 190 μm and 58.5 μm, respectively, with periods of $p_x = 296.4$ μm and $p_y = 475$ μm. The membrane metasurfaces were fabricated using conventional photolithography and deep reactive ion etching (DRIE) processes with a microscopic image shown in the inset of Fig. 2G. Transmission spectra measurements (Fig. 2G) demonstrate a transmission amplitude of over 96%, along with a full $2\pi$ phase coverage enabled by the two q-BIC resonances, consistent with simulations (see Methods). In addition, the q-BICs provide an alternative channel for boosting the $Q$ factor with enhanced local fields. An enhancement factor of 33 ($A_E=|E_{max}/E_0|$, where $E_{max}$ is the maximum local field amplitude and $E_0$ is the incident electric field amplitude) was achieved at $Q = 100$ in Fig. 2H. Precise control of quality factors while maintaining the extended Kerker effect is demonstrated in supplementary information section 3.

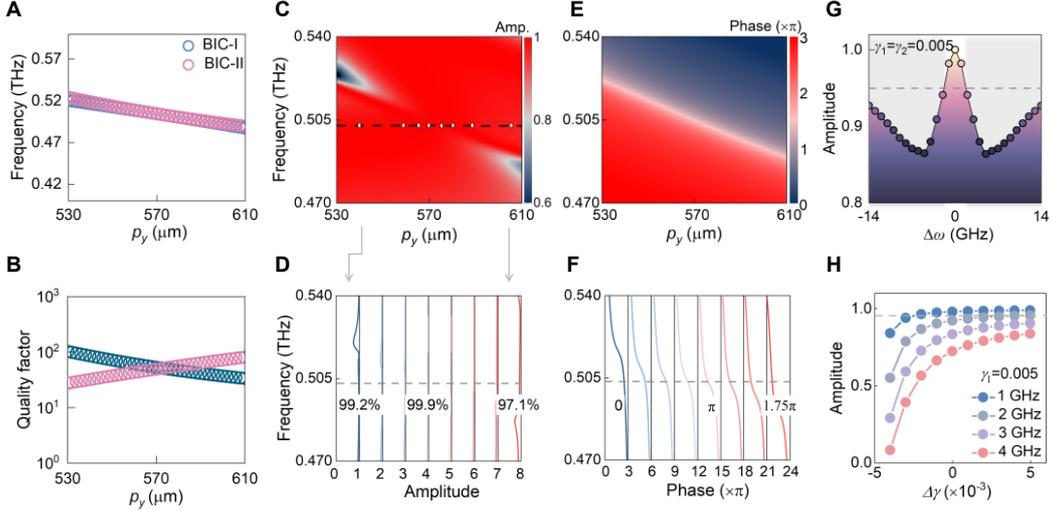

**Fig. 3 Extended Kerker effect with degeneracy of dual q-BICs in real space and searching for phase gradient unit cells with ideal efficiency.** (**A**), (**B**) Frequency and quality factor evolutions of the dual q-BICs in real space by merely tuning the period $p_y$ of unit cell along $y$ axis (corresponding to change of wavevector $k_y$). (**C**), (**E**) Look-up tables of transmission amplitude and phase versus parameter $p_y$. We plotted unwrapped phase that is accumulated with frequency, and total phase coverage therefore exceeds $2\pi$ in the band of interest. (**D**), (**F**) Transmission spectra of eight unit cells constructing a supercell with neighboring phase difference of $\pi/4$ ($p_y$ = 540 μm, 558 μm, 566 μm, 571 μm, 575.5 μm, 581 μm, 588 μm, and 606 μm). (**G**) Transmission amplitude variation versus frequency mismatch. The radiative losses of the two q-BICs are set equal, and the mismatch of central frequency from -14 GHz to 14 GHz (5.6% of central frequency) reveals substantial amplitude variation. (**H**) Transmission amplitude variation versus radiative loss mismatch. A smaller frequency mismatch between the dual q-BICs leads to a more robust transmission amplitude with respect to variations in radiative losses.

## Extended Kerker effect in parameter space

The extended degeneracy of dual q-BICs observed in momentum space is transformed into the extended Kerker effect by adjusting the $y$-axis period in parameter space. As revealed in Fig. 3A and 3B, the degeneracy is preserved over a large range of $p_y$ from 530 μm to 610 μm, with only a slight deviation in the $Q$ factors between the two q-BIC resonances. Lookup tables for amplitude and phase are summarized in Fig. 3C and 3E, generated by varying $p_y$ while keeping the remaining parameters fixed at $a$ = 246.4 μm, $b$ = 103.8 μm, $p_x$ = 385 μm, $\alpha$ = 10.7°, $h$ = 113.5 μm. As a representative example, eight unit cells supporting transparent transmission with a phase difference of $\pi/4$ at a central frequency of 0.5 THz are identified (Fig. 3D and 3F). Within the extended Kerker effect, an average transmission efficiency of 99.3% is achieved across the eight unit cells. The factors affecting the transmission coefficients were numerically analyzed involving deviation in central frequencies and losses between the two modes. In the absence of coupling between modes 1 and 2, the transmission coefficient of two typical Lorentzian resonances can be expressed as:

$$t = \left| 1 - \frac{\gamma_{1r}}{i(\omega-\omega_1)+\gamma_{1r}+\gamma_{1nr}} - \frac{\gamma_{2r}}{i(\omega-\omega_2)+\gamma_{2r}+\gamma_{2nr}} \right| \tag{1}$$

where $\gamma_r$ and $\gamma_{nr}$ represent radiative and non-radiative losses, respectively; and $\omega_1$ and $\omega_2$ are central frequencies of the two modes (see supplementary information section 1). The relative relationship between $\Delta\omega$ (frequency deviation) and $\Delta\gamma$ (loss difference) determines the resultant lineshape. Under the conventional Kerker's condition, the transmission coefficient rapidly drops below 95% (Fig. 3G) when $\Delta\omega/\omega = 0.2\%$ under moderate radiative losses of 0.005 ($\Delta\gamma = 0$ and $Q = 50$ at 0.5 THz), which is insufficient for achieving a full phase gradient coverage in parameter space (supplementary fig. S4). In contrast, greater robustness to $\Delta\gamma$ is observed at a smaller $\Delta\omega$ as indicated in Fig. 3H. This highlights the critical importance of mode degeneracy in frequency and the robustness against variations in $Q$ factors under the extended Kerker effect (supplementary fig. S5). These features ensure high transmission efficiency and broad phase gradient coverage across a wide range of design parameters.

With the exceptionally high transmission coefficient achieved for each unit cell, we fabricated a phase gradient metasurface as illustrated in Fig. 4A, where only $p_y$ is adjusted in unit cells. The design principles are detailed in supplementary information section 5. The deflection angle ($\theta$) is approximately estimated using the generalized Snell's law[4]:

$$\theta = \arcsin\left(\frac{\lambda_0 d\varphi}{2\pi dy}\right) \tag{2}$$

where $\lambda_0$ is the central wavelength, and $d\varphi$ and $dy$ are the phase shift and geometric distance between neighboring unit cells along the $y$-axis, respectively. The incidence electric field is $x$-polarized to minimize scattering caused by periodic disturbances, thereby maintaining the high local field enhancement. Using an angle-resolved terahertz time-domain spectroscopy, we directly captured the far-field time-domain signals of spatial distribution radiated from the metasurface as shown in Fig. 4B. Most energy is concentrated in the $0^{th}$ order diffraction and diminishes when deviating from 0°. At +7.5° (+$1^{st}$ order diffraction), a strong and long-lasting oscillation reappears, while no signal is captured at -7.5° (-$1^{st}$ order diffraction). A closer examination of the time-domain signals at the three angles is presented in Fig. 4C.

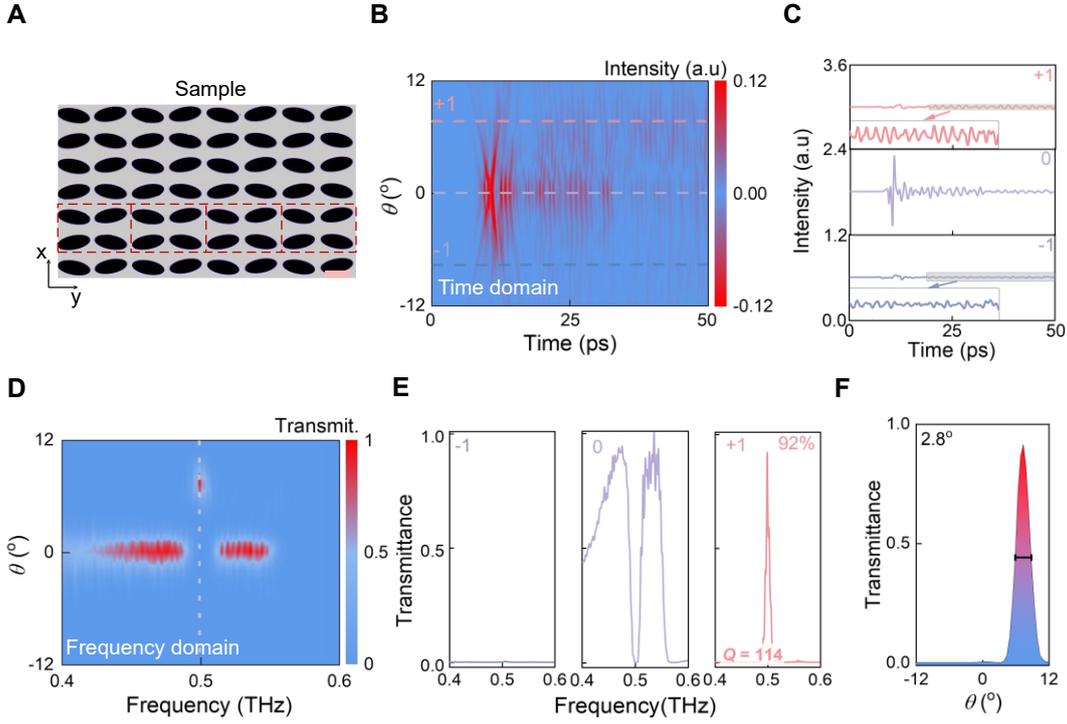

**Fig. 4 Experimental results of the membrane metasurfaces under extended Kerker effect.** (**A**) Optical microscopic image of fabricated free-standing sample with only varied period $p_y$ of unit cells. The microscopic image of the sample was obtained using an OLYMPUS BX53M optical microscope equipped with a ×5 objective lens. Scale bar, 200 μm. (**B**) Angle-resolved transmission spectra in time domain. Main pulse energy is concentrated in the zero-order diffraction while strong and long-lasting oscillations are observed in +1st order diffraction, as shown in the zoomed graph in (**C**). (**D**) Angle-resolved transmission spectra in frequency domain showing superior spectral and spatial selectivity at +1st order diffraction. (**E**) The zoomed-in transmission spectra at 0th and ±1st order diffraction angles. Complete darkness is experimentally observed at 0.5 THz in 0th order diffraction, and the energy is deflected toward +1st order diffraction with an absolute efficiency of 92% and quality factor of 114. (**F**) Spatial selectivity of +1st order diffraction with a divergent angle of 2.8°.

The performance of metasurface in the frequency domain was analyzed by Fourier transforming the time-domain signals. The measured angle- and frequency-resolved data are plotted in Fig. 4D. An exceptional spectral and spatial selectivity is observed in the +1st order diffraction. In the range of 0.4 to 0.6 THz, energy deflection toward the +1st order diffraction is nearly complete, with less than 0.2% energy measured in the -1st order diffraction and complete suppression (darkness) at the central frequency of the 0th order diffraction. Consequently, energy deflection toward the +1st order diffraction achieves an efficiency as high as 92%, experimentally realized at 7.5° (Fig. 4E, see Methods).

In the spatial domain, the divergent angle of the +1st order diffraction was measured to be 2.8° in the output domain, demonstrating excellent spatial selectivity of the deflected beam which originates from the narrow bandwidth empowered by q-BIC resonances (Fig. 4F)[11, 17]. Moreover, the exceptional spectral selectivity enabled by the q-BICs results in a $Q$ factor of 114 in the +1st order diffraction resonance, providing favorable conditions for low-threshold modulation. The combination of record-breaking efficiency, free-standing thin-film membrane, and outstanding spectral and spatial selectivity in the output domain in a single optoelectronic device is unprecedented (see supplementary table S2).

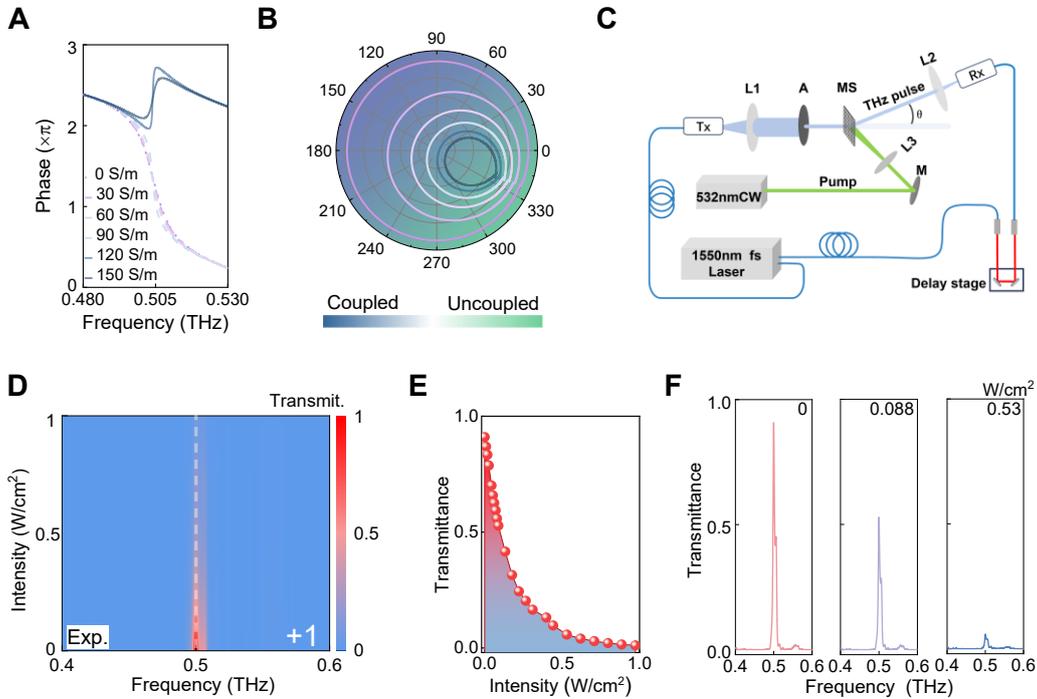

**Fig. 5 Experimental observation of low-threshold modulation.** (**A**), (**B**) Phase spectra and parametric polar plot of transmission in membrane metasurface by tuning nonradiative losses. Transition between underdamped and overdamped regimes occurs when the balance of losses is flipped. (**C**) Schematic diagram of experimental setup demonstrating the modulation of deflected beam. The incident terahertz beam is collimated with beam area of 0.5 cm², smaller than pump beam spot (532 nm, 1.13 cm²). (**D**) Measured modulation of transmission spectra versus pump intensity ranging from 0 W/cm² to 1 W/cm², recorded at +1st diffraction order. (**E**), (**F**) Zoom-in plot of modulation versus pump intensity. 94% modulation of the +1st order diffraction with pump intensity of 0.53 W/cm².

**Experimental verification of low-threshold modulation under extended Kerker effect**

We experimentally validated the modulation performance of the high $Q$ +1$^{st}$ order diffraction. According to Eq. 1, the degenerate q-BIC resonances enable a full $2\pi$ phase coverage in the underdamped regime when nonradiative losses are lower than radiative losses as illustrated in Fig. 5A. However, when the loss balance is reversed, a sudden transition occurs, reducing the phase coverage to less than $\pi$[48]. The transition is visualized through a parametric polar plot, illustrating the underdamped and overdamped regimes by observing whether the spectrum encircles the center point (Fig. 5B). This phase transition disrupts the extended Kerker effect within the supercell, eliminating the predesigned phase gradient and the consequent in-plane wavevector in the absence of $2\pi$ phase coverage. Concurrently, modifications in loss also deteriorate the transmission efficiency, reducing the diffraction efficiency.

The modulation performance was demonstrated using a reconfigured optical pump terahertz probe setup (OPTP, Fig. 5C). Photocarriers were injected into the silicon film using a continuous-wave (CW, 532 nm) laser. The pump beam spot (1.13 cm$^2$) was larger than the terahertz probe beam, ensuring uniform photoexcitation in the probe region. By gradually tuning the pump intensity, the captured deflected signals at +1$^{st}$ order diffraction dimmish (Fig. 5D and 5E). The deflected intensity was modulated from 92% to 5.8% at 0.5 THz with a pump intensity of only 0.53 W/cm$^2$ (Fig. 5F), corresponding to a modulation efficiency of 94%. As a comparison, less than 7% modulation was observed for an unpatterned silicon film under the identical conditions at pump intensity of 1 W/cm$^2$ (supplementary fig. S8). Modulation experiments using a pulsed laser (100 fs, 800 nm) were also conducted to validate the low-threshold performance (supplementary fig. S9). As a promising extension of BICs, further enhancement of the $Q$ factor of the deflection beam is feasible, which will improve spectral selectivity and reduce the modulation threshold even further.

**Discussion**

Metadevices based on nonlocal BICs present transformative potential for ultra-compact, planarized optical systems, allowing for functionalities such as spatial compression, augmented reality, and the generation of light bullets[49]. Realizing these capabilities hinges on overcoming key challenges in nonlocal metasurfaces, including efficiency, power consumption, and beamforming precision. In this study, we achieve a record-breaking 92% efficiency and excellent spectral and spatial selectivity of deflected beam in nonlocal q-BIC metasurfaces by implementing the extended Kerker effect. This was made possible by leveraging extended degeneracy and narrow resonance linewidth in a membrane metasurface. The selectivity in spectral and spatial domains is characterized by a 4 GHz bandwidth and a divergent angle of 2.8° for the deflected beam. Additionally, we demonstrated the low-threshold modulation facilitated by high-$Q$ resonances, further expanding the practical applicability of this approach. This strategy holds promise for advancing free-space beam navigation in next-generation wireless communication systems, including transmitters, processors, and relays. Moreover, the extended Kerker effect is scalable to shorter wavelengths, offering broad applicability across various photonic and optoelectronic platforms.

**Materials and Methods**

**Simulations**

Numerical simulations were conducted using commercially available software (COMSOL Multiphysics) with the RF module of the finite-element frequency-domain solver. Periodic boundary conditions were utilized for the unit cell, and perfectly matching layers (PML) were applied at the input and output ports. A non-dispersive refractive index of $n = 3.42$ was assigned to silicon in the unit cell. The modulation of losses was modeled by giving a finite value of conductivity to a 10 μm thick silicon film. Consistent with experiments, the quench of the 1$^{st}$ order diffraction intensity was realized by increasing the conductivity of silicon to 200 S/m.

**Fabrications**

Prior to fabrication, the silicon thin film was cleaned in an ultrasonic bath with acetone for 10 min and rinsed with isopropanol, followed by baking on a hot plate at 120 °C for 180 s. Then, 800 nm thick $SiO_2$ was deposited on the silicon as an etching hard mask using the plasma-enhanced chemical vapor deposition (PECVD-PlasmaPro 80). Photoresist (RZJ 304.50) was spin-coated on the sample at a speed of 5000 r/min for 30 s with a thickness of 2 μm. The sample was then baked on a hot plate (100 °C, 180 s). Conventional UV photolithography (SUSS-MA6) was used to transfer the predesigned pattern on the photoresist, and then the sample was developed with the RZX3038 developer for ~ 30 s. The patterned sample was baked on a hot plate at 120 °C for 90 s. The $SiO_2$ layer was then etched with reactive ion etching (RIE), forming a hard mask for the subsequent deep reactive ion etching (DRIE, Oxford Estrelas) process of the silicon film. Finally, the $SiO_2$ residue was removed by immersing the sample in BOE solution.

**Measurements**

In the reconfigured angle-resolved measurement system (Fig. 5C), we maintained fixed positions for the terahertz transmitter (Tx), collimating lens L1 ($f$ = 50 mm), precision pinhole aperture (A), and sample stage, while mounting the collecting lens L2 ($f$ = 50 mm) and receiving antenna (Rx) on a rotary translation stage for precise angular detection ($\theta$ = ±90° with 0.5° resolution). The conductivity of the metasurface was dynamically modulated via optical excitation using a 532 nm continuous-wave laser with adjustable power (0-2.4 W). The incident terahertz beam, collimated to a spot size of 0.5 cm$^2$, was intentionally kept smaller than the pump laser illumination area (1.13 cm$^2$) to ensure uniform photoexcitation across the probed region. The raw time-domain signals were Fourier-transformed to obtain frequency-domain spectra, which were then normalized against the reference signals measured in free space to eliminate the background caused by the system response. Here, $E_s(\omega)$ and $E_r(\omega)$ represent the complex spectra of the sample and the air reference, respectively. Transmission amplitude is defined as $t = |E_s(\omega)/E_r(\omega)|$ and phase is obtained by $p = \arg(E_s(\omega)) - \arg(E_r(\omega))$. Transmittance is characterized by $t^2$ indicating the intensity information. The quality factor can be extracted by using the Fano lineshape equation[50]:

$$\left|t^2\right| = \left|a_1 + ja_2 + \frac{b}{\omega - \omega_0 + j\gamma_{\text{tot}}}\right|^2 \tag{3}$$

where $a_1$ and $a_2$ are real constants, $\gamma_{tot}$ is the total loss rate, and $\omega_0$ is the central frequency of resonance. $Q_{tot}$ was determined by $Q_{tot} = |\omega_0/2\gamma_{tot}|$.

**Data Availability**

The data are available from the corresponding author upon reasonable request.

**Acknowledgments**

This work was supported by the National Natural Science Foundation of China (Award No.: 62335011, 62175099, 62475134), National Key R&D Program of China (2024YFA1410100), Guangdong Basic and Applied Basic Research Foundation (Award No.: 2023A1515011085, 2025A1515010064), Guangdong Provincial Quantum Science Strategic Initiative (Award No.: 2023A1515011085), Shenzhen Science and Technology Program (Award No.: JCYJ20241202125300002, JCYJ20230807093617036), and high level of special funds from Southern University of Science and Technology (G030230001, G03034K004). The authors acknowledge the assistance of SUSTech Core Research Facilities, Rui Zhang, and Yao Wang.

**Author Contributions**

Supervision: L. C. and Y. Z. Fabrication: Y. Z. and J. F. Investigation, simulation, data analysis, measurement and original draft: J. F. Writing – review & editing: J. F., Z. X., G. X., J. C., H. X., Y. Z. and L. C.

**Competing interests**

The authors declare no competing interests.